\documentclass[reprint, amsmath,amssymb,aps,prl]{revtex4-1}

\usepackage{graphicx}
\usepackage{dcolumn}
\usepackage{bm}
\begin{document}

\preprint{APS/123-QED}

\title{Robust manipulation of light using topologically protected plasmonic modes}

\author{Chenxu Liu}
\affiliation{Department of Physics and Astronomy, University of Pittsburgh}

\author{M.V. Gurudev Dutt}
\affiliation{Department of Physics and Astronomy, University of Pittsburgh}
\affiliation{Pittsburgh Quantum Institute, University of Pittsburgh}

\author{David Pekker}
\affiliation{Department of Physics and Astronomy, University of Pittsburgh}
\affiliation{Pittsburgh Quantum Institute, University of Pittsburgh}

\date{\today}

\begin{abstract}
We propose a topological plasmonic crystal structure composed of an array of parallel nanowires with unequal spacing. In the paraxial approximation, the Helmholtz equation that describes the propagation of light along the nanowires maps onto the Schr\"{o}dinger equation of the Su-Schrieffer-Heeger (SSH) model. Using full three-dimensional finite difference time domain solution of the Maxwell equations we demonstrate the existence of topological defect modes, with sub-wavelength localization, bound to kinks of the plasmonic crystal.  Furthermore, we show that by manipulating kinks we can construct spatial mode filters, that couple bulk modes to topological defect modes, and topological beam-splitters that couple two topological defect modes. Finally, we show that the structures are robust to fabrication errors with inverse length-scale smaller than the topological band gap. 
\end{abstract}

\maketitle

\section{\label{sec:level1}Introduction}
The interface between two electronic materials with topologically distinct band structures necessarily supports a topologically protected mode~\cite{KaneMele2005,Bernevig2006,FuKane2007}. The robustness of this mode arises from global symmetries like parity, space inversion, and time reversal~\cite{Schnyder2008, Kitaev2009}. In electronic systems, the topology of the electronic band structures can be controlled relatively easily by either using materials properties (e.g. picking a suitable semiconductor~\cite{Hasan2010RMP}) or by materials engineering (e.g, by making hybrid structures like quantum wells out of suitable materials~\cite{Konig2007Science}). The combination of robustness of topological edge modes and the relative ease of fabrication has resulted in an explosion of interest in these systems with applications ranging from decoherence-free quantum state manipulation~\cite{Nayak2008RMP} to electronic device design~\cite{Xiu2011,Nakayama2012,Mellnik2014Nature}. 

Following the seminal work of Ref.~\cite{Haldane2008}, it was realized that topological band structures can also exist in photonic systems. These ideas have been explored in a number of theoretical proposals~\cite{Khanikaev2013,Lu2013,Rechtsman2013,Skirlo2014,Lu2016}. They have also been realized experimentally in the following photonic crystal systems: gyromagnetic photonic crystal at microwave frequencies~\cite{Wang2009}, coupled whispering gallery mode optical resonators~\cite{Hafezi2013}, and spiral optical waveguides~\cite{Rechtsman2013}. For an in depth review of the recent progress and challenges, see Ref.~\cite{LuReview2014}.

In this work, we propose a two dimensional plasmonic crystal structure that supports topologically protected defect modes for light traveling near to the axial direction, similar to Ref.~\cite{Rechtsman2013}. These modes are protected by the sublattice symmetry of the Su-Schrieffer-Heeger (SSH) model~\cite{HeegerKivelsonSchriefferSuRMP1988,Schnyder2008}, which belongs to the AIII-class of Ref.~\cite{Schnyder2008}. Our proposal opens up the domain of metal nanostructures for studying topological light, as compared to previous efforts that focused on dielectric structures. Surface plasmon-polariton modes on metal nanowires in plasmonic crystals have the advantage of being (a) broadband, (b) relatively easy to manufacture, and (c) capable of sub-wavelength localization of light.

The proposed setup allows us not only to guide light but also to robustly manipulate it by shifting the edge modes, as a function of axial position, inside the structure. Using full 3D finite-difference-time-domain (FDTD) solutions of the Maxwell equations, we demonstrate a beam splitter  and spatial mode filter (that couples light from a pair of bulk modes to a pair of edge modes). Further, using the 3D FDTD calculations we show that the structures are tolerant to manufacturing defects with transverse length-scales comparable to inter-nanowire spacing, as long as the axial length-scales are larger than the inverse topological band gap. 

\section{Guiding light using topological defect modes}
Consider a system that is almost translationally invariant along the axial, i.e. $z$-direction. Our goal is to describe paraxial modes, time-harmonic electromagnetic waves that propagate at small angles to the z-axis. For paraxial TM-modes it is natural to focus on the transverse components of the electric field~\cite{Huang1992,Huang1993} and to separate out the fast oscillating part,  $E_x=\psi_x(x,y,z) e^{i\beta_0 z}$ and $E_y=\psi_y(x,y,z) e^{i\beta_0 z}$, where $\beta_0=\omega/c$.  The propagation of paraxial TM-modes is governed by the paraxial Schr\"{o}dinger equation (see supplement for details)
\begin{equation}
\left[-\frac{c}{2\beta_0}\nabla_\perp^2
+ V\right]
\left(\begin{array}{c}
\psi_x\\
\psi_y
\end{array}\right)=i\, c\, \partial_z
\left(\begin{array}{c}
\psi_x\\
\psi_y
\end{array}\right),
\label{eq:ParaApp}
\end{equation}
where, $\nabla_\perp^2 = \partial_x^2+\partial_y^2$, $V(x,y,z)$ describes the position of the metal nanowires, and we have neglected $\partial_z^2 \psi_{\{x,y\}}$ as per the paraxial approximation. Eq.~\eqref{eq:ParaApp} has the form of the two-dimensional time-dependent Schr\"{o}dinger equation with $c\,\partial_z \rightarrow  \partial_t$, $\psi_{\{x,y\}}$ -- the wave function, and $H \equiv -(c/2\beta_0)\nabla_\perp^2+V(x,y,z)$ -- the Hamiltonian operator. Thus, stationary states of $H$ become the plasmon modes that propagate along the $z$-direction and the eigenenergies $\varepsilon$ of $H$ become the z-wavenumbers $\beta_z=\beta_0-\varepsilon/c$. Ref.~\cite{Rechtsman2013} showed that this analogy can be used to gain intuition about the topological structure of electromagnetic waves by mapping solutions of the Schr\"{o}dinger equation with non-trivial topology onto the Helmholtz equation.

\begin{figure}[ht]
\includegraphics[width=3.4 in]{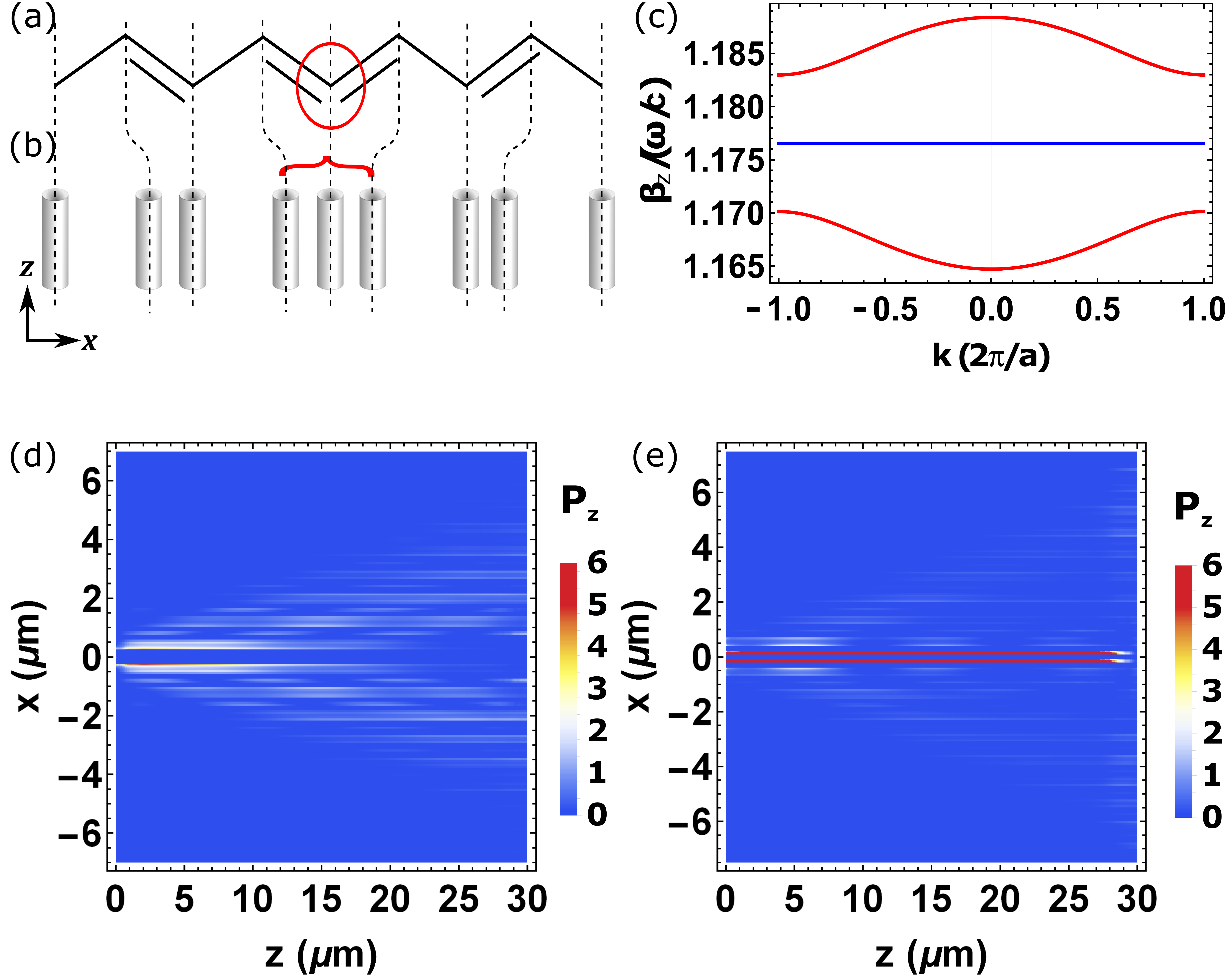}
\caption{\small{(a) Single-double bond pattern in a polyacetylene chain with a kink defect. (b) Analogous nanowire array with staggered spacing and a kink defect. (c) Staggered nanowire spacing leads to a band gap for the bulk modes (red), the topological defect mode appears in the middle of the band gap (blue). (d) Spreading of light in a nanowire array without a kink. (e) Light guidance by a topological defect mode in a nanowire array with a kink. Parameters used: material relative permittivity $\epsilon = -45.83 + 2.49 \times 10^{-9} i$, diameter $200\,\text{nm}$, center-to center spacing $300$ and $500\,\text{nm}$, $\lambda=1\,\mu\text{m}$. }}
\label{fig:1}
\end{figure}

Let us now consider the SSH model, which was originally used for studying electrons in polyacetylene. The backbone of polyacetylene is a chain of carbon atoms with staggered single and double bonds~\footnote{Here, we assume that the pattern of single and double bonds does not have intrinsic dynamics, but is instead a prescribed function of time.}, schematically depicted in Fig.~\ref{fig:1}(a).  The discretized version of the Hamiltonian describing the hopping of spinless electrons along the backbone of polyacetylene is 
\begin{equation}
H_{\text{SSH}}=-\sum_i 
t_{i,i+1}
\left(c_i^\dagger c_{i+1}+c_{i+1}^\dagger c_i\right)+v_i c_i^\dagger c_i
\label{eq:ssh}
\end{equation}
where the operator $c_i$ ($c_i^\dagger$) annihilates (creates) an electron on the $i$-th carbon atom, the hopping matrix element $t_{i,i+1}=t_1 (t_2)$ if the bond between sites $i$ and $i+1$ is a single (double) bond, and $v_i$ is the on-site energy. The SSH model band structure has a band gap (see Fig.~\ref{fig:1}(c)) as long as the tunneling matrix elements $t_1$ and $t_2$ are unequal. A kink in the single-bond double-bond pattern of the chain, as depicted in Fig.~\ref{fig:1}(a), is a topological defect: it cannot be removed by local manipulations of the chain and can only be eliminated by merging it with another kink. Moreover, each kink must host a mid-gap state, a topological defect mode, that is localized in the vicinity of the kink. 

We now investigate the optical equivalent of the SSH kink modes in plasmonic crystals. Consider a plasmonic crystal that consists of an array of parallel nanowires with staggered spacing as depicted in Fig.~\ref{fig:1}(b). The Helmholtz-Schrodinger analogy tells us that for each eigenmode of the SSH Hamiltonian Eq.~\eqref{eq:ssh} there is an equivalent electromagnetic mode in the plasmonic crystal. We note that the fermionic commutation relations do not play a role here as we are considering the non-interacting case; hence, we can replace the operator $c_i^\dagger$ that creates an electron in a carbon atom atomic orbital by the operator $b_i^\dagger$ that create a surface plasmon on the i-th nanowire~\footnote{We present the details of the connection between the continuous and discrete Hamiltonians in the supplement.}. Thus the band gap in the electronic system maps onto a $\beta_z$ gap in the plasmonic system. Moroever, electronic states that are localized on kinks (and appear inside the band gap) map directly onto guided plasmonic states that propagate along the kinks in the z-direction (and appear in the $\beta_z$ gap, see Fig.~\ref{fig:1}(c)).

Now consider injecting a spatially truncated plane wave into a plasmonic crystal with staggered spacing but no kinks. This is equivalent to injecting an electron into polyacetylene using a local probe like an STM tip. Because the electron is being injected locally, it overlaps many k-modes, and hence the electron wave-packet will spatially spread out as time advances. Similarly, the plasmonic wave packet will expand in the transverse direction as it advances along the $z$-direction. To illustrate this expansion we introduce the normalized Poynting vector $P_z$ in the x-z plane that cuts through the middle of the nanowire array 
\begin{equation}
P_z(x,y=0,z) = \frac{\vec{S}(x,y=0,z)\cdot \hat{z} }{\int dx\, dy \, \vec{S}(x,y,z)\cdot \hat{z}},
\end{equation}
where $\vec{S}$ is the non-normalized Poynting vector and $\hat{z}$ is the unit vector along the z-direction. Fig.~\ref{fig:1}(d) shows the spreading out of a plasmonic wave-packet obtained using 3D FDTD simulation of the plasmonic crystal. Completing the analogy, the group velocity of the electron in polyacetylene corresponds to the opening angle of the light cone in the plasmonic crystal. 

The introduction of a topological defect into the plasmonic crystal (see Fig.~\ref{fig:1}(b)) gives rise to a localized mode. We expect that light injected in the vicinity of the topological defect couples both to the localized defect mode as well as to the bulk modes~\cite{Rechtsman2013}. We plot the results of injecting a truncated plane wave into a plasmonic crystal with a topological defect in Fig.~\ref{fig:1}(e). In accord with our expectations, we observe that light coupled into the bulk modes forms a diffracting fan, while light coupled into the topological mode propagates without spreading transversely. 

We comment that light guided by a topological defect mode is spatially concentrated. As a figure of merit, we consider the quantity $\lambda^2 P_z$ which measures how much the light is squeezed spatially as compared with diffraction limited optics ($\lambda^2 P_z \approx 1$ at diffraction limit). Plasmonic confinement of light in the topologically guided mode of the structure depicted in Fig.~\ref{fig:1}(e) results in $\lambda^2 P_z \approx 16.6$.

\section{Manipulation of light using topological defect modes} 
In the electronic system, there are two well-established operations for manipulating topological defect modes: (i) shifting the position of a kink causes the associated topological defect mode to be carried along with the kink, and (ii) pairs of kinks can be nucleated and pulled apart, causing two of the bulk modes to be turned into topological defect modes.  In this section, we explicitly apply these operations to achieve topological manipulation of light in the plasmonic crystal of metal nanowires. 

In the Helmholtz-Schr\"{o}dinger correspondence, $\partial_t$ is mapped to $c \partial_z$, so that time dependent manipulation in the electronic system is mapped onto axial dependent manipulation in the plasmonic crystal. For instance, the time-dependent shifting of kinks in the electronic picture is mapped to the shifting of the kinks as a function of the axial position $z$ in the plasmonic crystal, which we achieve through the variation of the nanowire spacing as a function of $z$. 

\begin{figure}[ht]
\includegraphics[width=3.4 in]{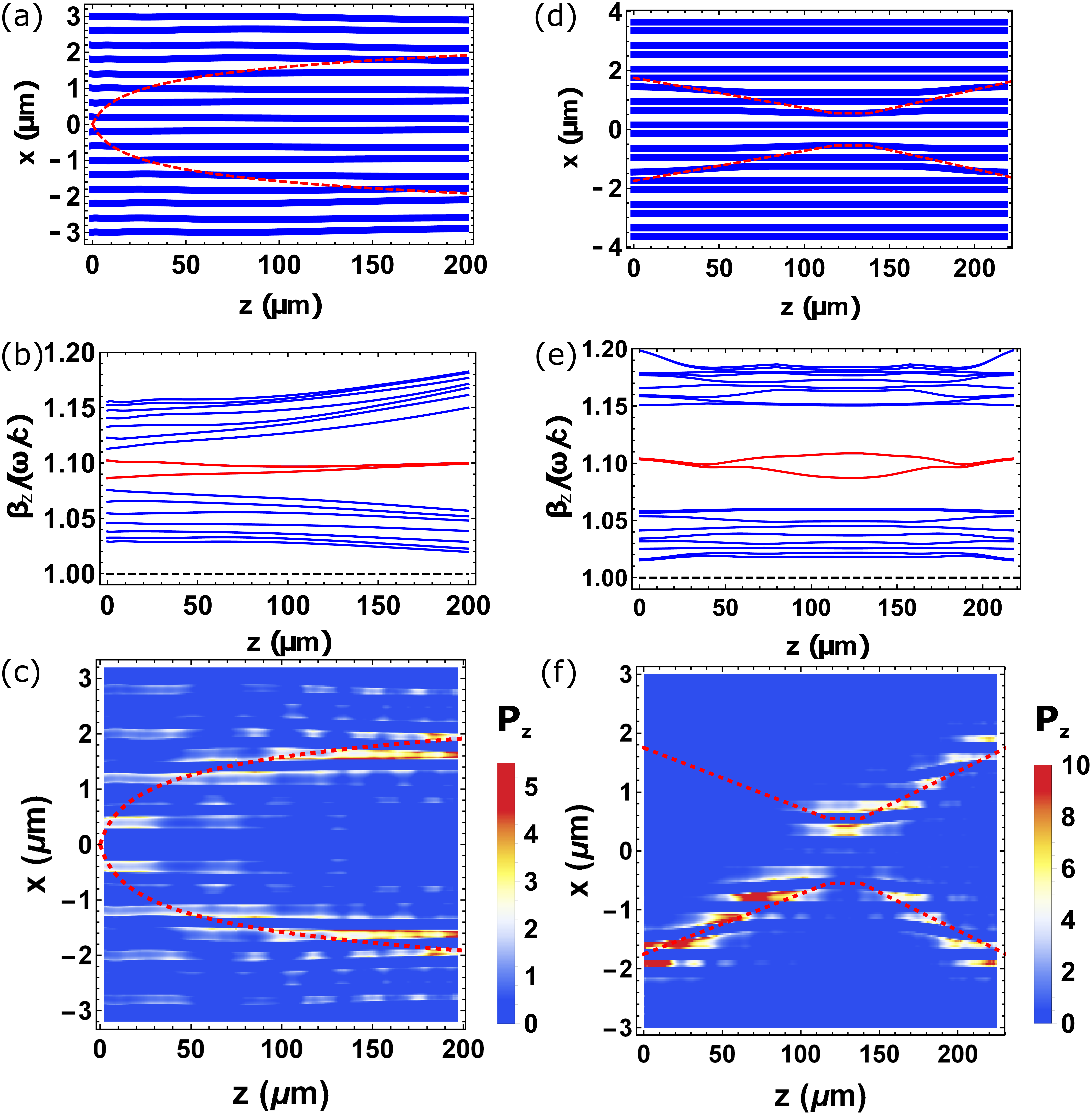}
\caption{\small{(a,d) Nanowire array geometry for a spatial mode filter (a) and a beam splitter (d). Blue lines indicate the positions of the nanowires and red dashed lines the position of the kinks. (b,e) Spectral flow of bulk modes (blue) and the topological defect modes (red) as a function of $z$ for the spatial mode filter (b) and the beam splitter (e). (c,f) Light propagation through the spatial mode filter (c) and the beam splitter (f), red dashed lines indicate the position of the kinks (see text). 
Parameters used are same as Fig.~\ref{fig:1}. }}
\label{fig:2}
\end{figure}

\subsection{Topological spatial mode filter}
Nucleating two kinks in the middle of the SSH chain and adiabatically moving them apart results in a spectral flow in which a pair of delocalized electronic states, one from the upper bulk band and one from the lower, are adiabatically transformed into the two mid-gap states spatially localized on the kinks. We take advantage of this spectral flow to perform spatial mode filtering of light using a plasmonic crystal.

Specifically, we design a plasmonic crystal with a pair of kinks that are created inside the array and shifted apart as a function of $z$, see Fig.~\ref{fig:2}(a). Two of the bulk modes at the input side ($z=0$) of the array are mapped into the two topological defect modes on the output side ($z=200\,\mu\text{m}$) by the spectral flow, see Fig.~\ref{fig:2}(b). If one of these two bulk modes is injected into the array, after propagation, the maximum of the light intensity on the output side will be strongly localized around the kinks. To verify this behavior, we perform full 3D FDTD simulations of the structure depicted in Fig.~\ref{fig:2}(a), and plot the results in Fig.~\ref{fig:2}(c). We observe that the majority of the light flux is indeed guided into the topologcial defect modes. On the other hand, if we inject any other bulk mode on the input side it will be rejected by the mode filter and the output light flux in the vicinity of the kinks will be small (see supplement).

\begin{figure}[ht]
\includegraphics[width=3.4in]{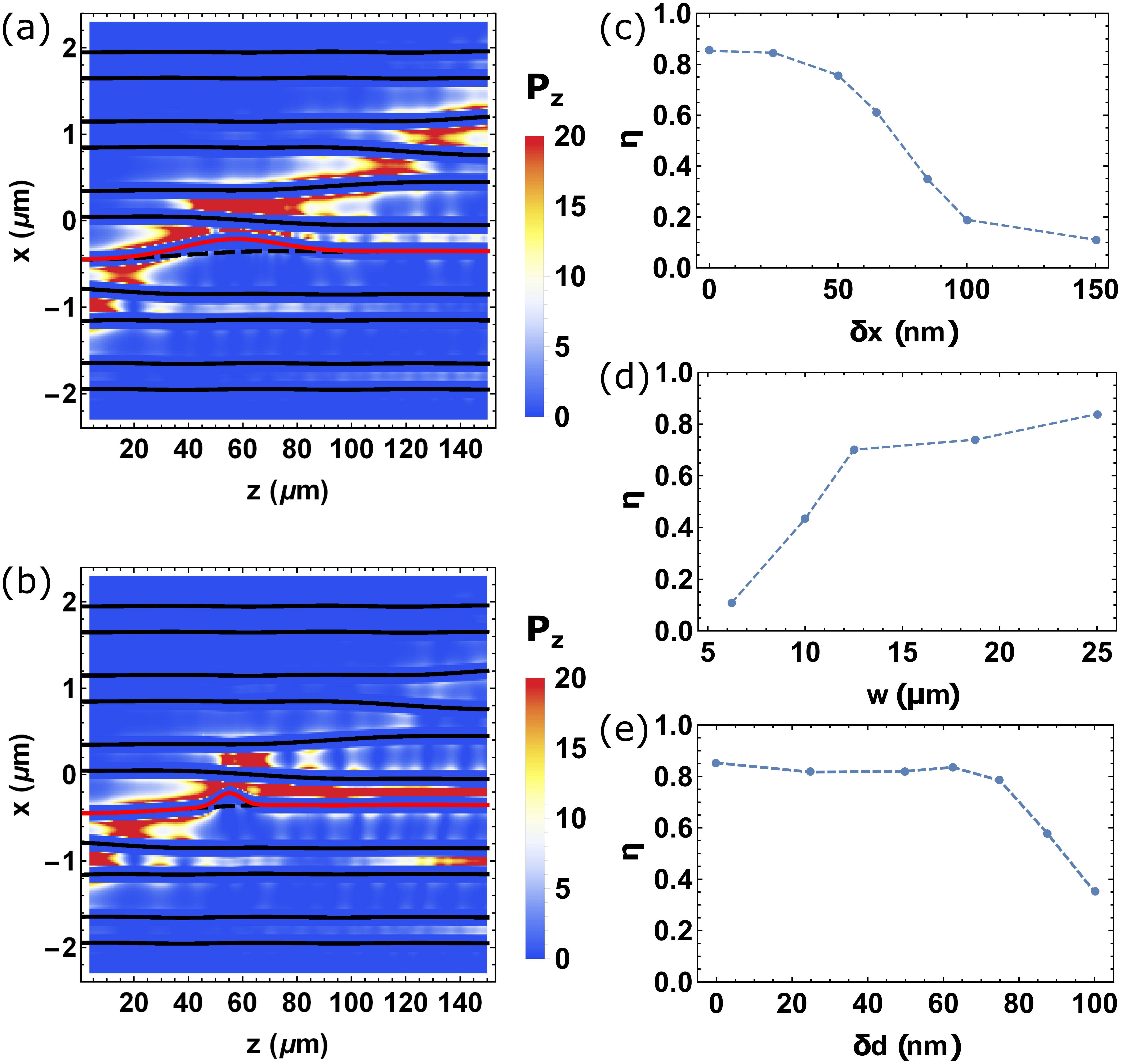}
\centering
\caption{\small{(a,b) Topological defect mode propagation in a nanowire array with a perturbation. The red nanowire is shifted away from its original position indicated by the black dashed line (see text for details). (c,d) Fraction of flux retained in the topological defect mode $\eta$ in a structure with a shifted nanowire. (c) $\eta$ as a function of the displacement parameter $\delta x$ (smoothness parameter is fixed at $w=6.25 \, \mu$m). (d) $\eta$ as a function of $w$ at fixed $\delta x=150\,\text{nm}$. (e) Similar to (c), except the nanowire diameter parameter $\delta d$ is varied and $w=25 \, \mu$m is fixed. Parameters used: same as Fig.~1. }}
\label{fig:3}
\end{figure}

\subsection{Beam splitter with topological defect modes}
Consider the geometry of a nanowire array depicted in Fig.~\ref{fig:2}(d): two kinks that are well separated spatially on the input and output sides of the structure are brought close together in the middle of the structure. The propagation of the two topological defect modes can be described by the Landau-Zener Hamiltonian, in which the coupling is controlled by the spatial separation of the kinks. The spectral flow of the array along the $z$ direction is shown in Fig.~\ref{fig:2}(e). When the kinks are well separated, the topological defect modes are non-interacting, and hence both are in the middle of the gap. As the kinks are moved closer, the topological defect modes begin to interact and the degeneracy is broken. By controlling the interaction strength and the length of the interaction region it is possible to construct a 50-50 beam splitter. We perform 3D FDTD simulation of the beam splitter in which we inject light into one of the defect modes on the input side and observe an equal superposition of light in the two defect modes on the output side (see Fig.~\ref{fig:2}(f)).

\section{Robustness against perturbations in wire placement and diameter}
Consider a structure in which a topological defect mode is guided by a kink that is being shifted as a function of $z$ (see Fig.~\ref{fig:3}). To test the robustness of the topological defect modes to perturbations we measure the amount of light that leaks from the topological defect mode into the bulk modes due to perturbations in the plasmonic nanowire structure. This is a particularly stringent test of robustness, as perturbations near the kink can affect the structure of the defect mode directly as well as how the defect mode hops from nanowire-to-nanowire as the kink is shifted. As it is not possible to test robustness to all perturbations, we focus on two specific types that are likely to occur in experimental situations: meander of a single nanowire and distortions in the diameter of a single nanowire.

In Fig.~\ref{fig:3}(a,b), we compare the results of 3D FDTD simulation for two test structures in which we displace one of the nanowires by the addition of a meander $M(z) = \delta x\, e^{-(z-z_0)^2/w^2}$ with $\delta x=150$nm. In Fig.~\ref{fig:3}(a) the meander is smooth with $w = 25\, \mu$m, while in Fig.~\ref{fig:3}(b) the meander is abrupt with $w = 6.25\, \mu$m. The positions of unperturbed wires are indicated with black lines, the original position of the perturbed wire is indicated with black dashed line and it's new position with the red line. The color scale shows $P_z$ for light that is injected into the topological defect mode on the left side of the structure. We observe that if the meander is sufficiently smooth the topological mode remains guided along the domain wall [Fig.~\ref{fig:3}(a)]. However, if the meander is too abrupt there is significant leakage of light into the bulk modes [Fig.~\ref{fig:3}(b)]. 

A perturbation can break down topological protection in one of two ways -- either by breaking adiabaticity (i.e. an abrupt meander) or by breaching the topological gap (i.e. a large magnitude meander). In practice, as we close the gap between the topological defect mode and the continuum spectrum $\Delta \beta \rightarrow 0$, the adiabaticity condition $\partial_z \Delta\beta<< \Delta\beta^2$ is always violated first due to the diverging denominator. To quantify the amount of leakage caused by a perturbation we introduce the quantity $\eta$
\begin{align}
\eta=\frac{1}{2}\int dx\, dy\, (\vec{E}_\alpha \times \vec{H} + \vec{E}\times \vec{H}_\alpha)\cdot \hat{z},
\end{align}
where $\vec{E}_\alpha$ and $\vec{H}_\alpha$ are the electric and magnetic fields that correspond to the guided mode ($\eta=1$ perfect guidance, $\eta=0$ complete leakage). In Fig.~\ref{fig:3}(c) we plot $\eta$ as a function of the displacement parameter $\delta x$. For $\delta x \lesssim 50 \, \text{nm}$, $\eta$ remains large indicating robustness to the perturbation. The robustness is suddenly lost for $\delta x \gtrsim 50 \, \text{nm}$ when adiabaticity is lost as the topological mode begins to merge with the continuum spectrum (see supplement for details). A complementary picture of robustness to smooth perturbations and the loss of adiabaticity for abrupt perturbations emerges when we tune the the smoothness parameter $w$ [see Fig.~\ref{fig:3}(d)]. 

Perturbations of the nanowire diameter, as opposed to  position, have qualitatively similar effect on the propagation of the topological defect mode. We investigate a series of structures similar to the one depicted in Fig.~\ref{fig:3}(a), but instead of shifting the red wire, we modify its diameter $D(z)=D_0-\delta d\, e^{-(z-z_0)^2/w^2}$, where $D_0=200\,\text{nm}$ is the unperturbed diameter. We plot $\eta$ as a function of $\delta d$ in Fig.~\ref{fig:3}(e). Similar to Fig.~\ref{fig:3}(c), we observe robustness to small perturbations followed by a sharp drop in $\eta$ associated with the loss of adiabaticity. 

\section{\label{sec:level4}Outlook}
One of the key issue of plasmonic devices is the absorption of light due to finite optical conductivity in metals and hence a small but finite imaginary part of the dielectric constant. Our calculations show that light manipulation using topological plasmonics requires the propagation of light over $\sim 50$ wavelengths. To investigate losses in realistic structures, we have computed the complex wavenumbers $\beta_z$ for the topological defect mode in structures made from single-crystal gold nanowires. We find that $\lambda\sim 1.56 \mu\text{m}$ (in vacuum) is a sweet spot for our structures, with topological defect mode having a decay length of $\sim 100 \mu\text{m}$ (see supplement). From the experimental perspective, these are appealing lengths scales due to the availability of telecommunications diode lasers and the ease of nano-fabrication.

\bibliography{ref}

\section*{Online Supplement}

\subsection{The paraxial Schr\"{o}dinger equations}
The propagation of electromagnetic waves is governed by the Helmholtz equation
\begin{equation}
\left(\nabla^2 + \frac{\omega^2}{c^2} \epsilon \right) 
\left\{ 
\begin{array}{c}
\vec{E}\\
\vec{B}
\end{array}
\right\}=
-\left\{
\begin{array}{c}
\nabla \left( \epsilon^{-1}\,{\nabla \epsilon \cdot \vec{E}} \right)\\
\epsilon^{-1}\,{\nabla \epsilon}\times \left(\nabla \times \vec{B} \right)
\end{array}
\right\}
\label{eq:helmholtz}
\end{equation}
where $\epsilon(x,y,z)$ is the position-dependent relative permittivity, $\vec{E}(x,y,z)$, $\vec{B}(x,y,z)$ are the electric and magnetic field components, and the right hand side encodes the boundary conditions at the metal-air interface. For the case of dielectric waveguides, the dielectric constant tends to vary gently, and hence  the terms on the right hand side of Eq.~\ref{eq:helmholtz} can be neglected. Therefore it is natural to obtain the paraxial Sch\"{o}dinger equation from the Helmholtz equation for $E_z$. For metallic structures variations of $\epsilon$ cannot be neglected. However, for  structures that are translationally invariant in $z$-direction, $\nabla \epsilon$ only has transverse components. Hence, it is natural to focus on the transverse components $E_x$ and $E_y$, as the Helmholtz equations for those components have closed form.  For TM-modes propagating at small angles to the $z$-axis it is natural to implement the paraxial approximation, which leads us to the paraxial Schr\"{o}dinger equation, Eq.~\ref{eq:ParaApp}. The potential energy operator, that appears in Eq.~\ref{eq:ParaApp}, is given by 
\begin{align}
V(x,y,z)=
\frac{\omega}{2} (1-\epsilon) +
c\left(
\begin{array}{cc}
\nabla_x\left(\frac{\nabla_x \epsilon}{\epsilon}\right) & 
\nabla_x\left(\frac{\nabla_y \epsilon}{\epsilon}\right)\\
\nabla_y\left(\frac{\nabla_x \epsilon}{\epsilon}\right) & 
\nabla_y\left(\frac{\nabla_y \epsilon}{\epsilon}\right)
\end{array}
\right), \nonumber
\end{align}
where the second term accounts for the boundary conditions at the metal-air interface. 

\subsection{Connecting the continuous and discrete Helmholtz equations}
In this supplement, we connect the continuous description of electromagnetic waves with the discrete description of the SSH model, i.e. Eq.~\eqref{eq:ParaApp} and Eq.~\eqref{eq:ssh} of the main text. We note that this connection is only precise in the limit of weak coupling (i.e. when the distance between nanowires is sufficiently large compared to the wavelength of light). In the strong coupling limit the topological properties of the Helmholtz equation remain intact but the tight-binding model can no longer be used to accurately describe light propagation. Consequently we use full 3D FDTD solutions of the Helmholtz equation throughout the main text. 

Our strategy to make the connection is to (1) describe the plasmon ``self-energy" by modeling a single nanowire, and (2) describe the plasmon hopping by modeling two nanowires. We begin by considering a single nanowire of the type that makes up the plasmonic crystal. The single nanowire has a well defined plasmon mode, i.e. a radially symmetric solution of Eq.~\eqref{eq:ParaApp} of the form $\psi(x,y,z)=\psi_1(x,y) e^{i \beta_1 z}$. We can capture the plasmon self-energy $\beta_1$, by setting $v_i=c (\beta_0-\beta_1)$ in Eq.~\eqref{eq:ssh}. 

\begin{figure}[ht]
\includegraphics[width=3.4 in]{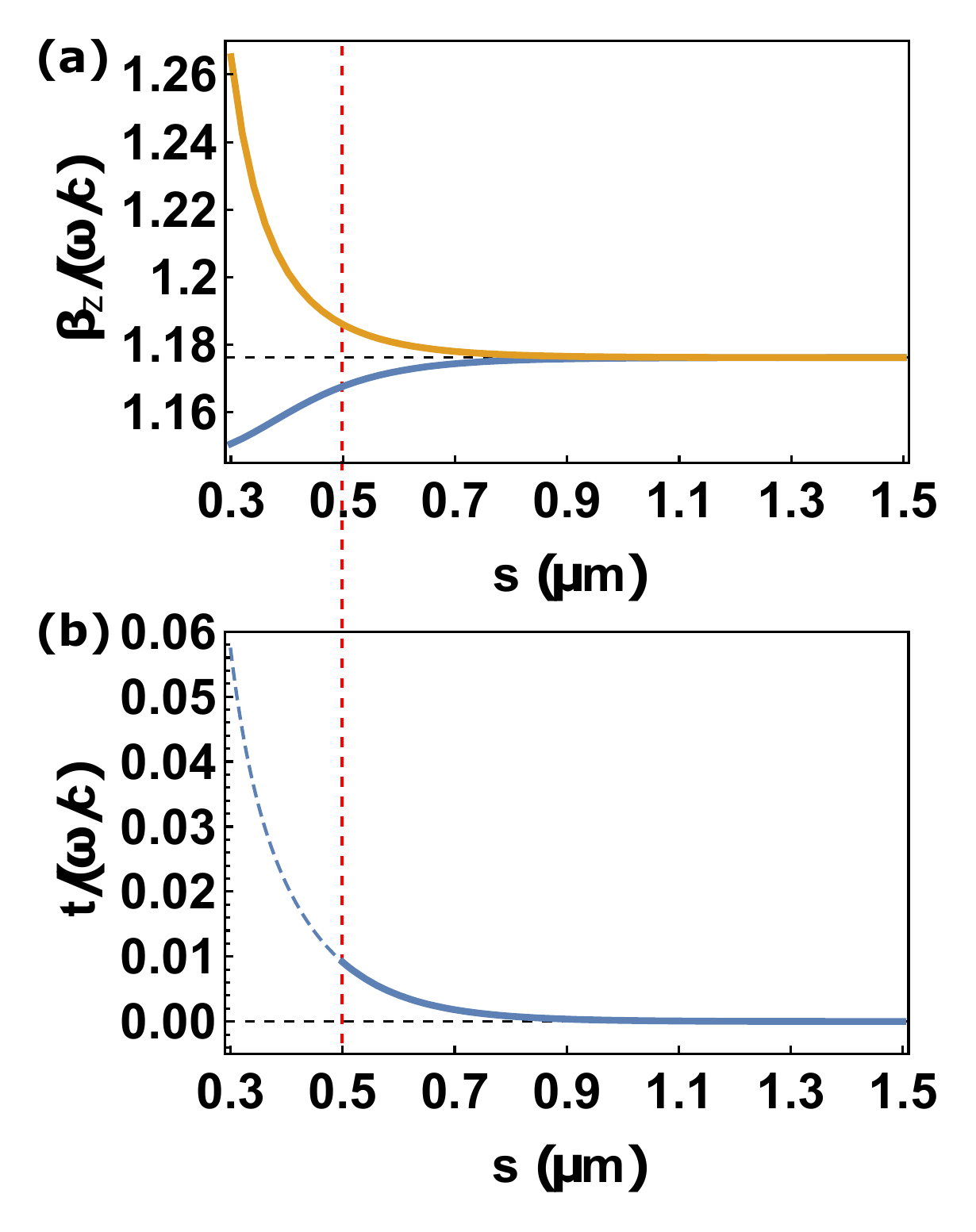}
\caption{\small{(a) The splitting of the symmetric and antisymmetric plasmon modes (for $\lambda=1 \,\mu \text{m}$ in vacuum) in a system of two parallel silver nanowires with $100$ nm radius, as a function of the nanowire center-to-center separation $s$. The black dashed line indicates the single nanowire eigenvalue $\beta_1$. (b) Extracted tight-binding parameter $t$ as a function of spacing between two nanowires. The tight-binding model starts to break down when the spacing between the two nanowire goes below $0.5 \, \mu \text{m}$ (indicated by the dashed red line).}}
\label{fig:sym_antisym}
\end{figure}

Next, we consider the plasmon modes of a system of two parallel nanowires separated by distance $s$. The plasmon spectrum is now composed of two modes $\psi_\pm$ with eigenvalues $\beta_\pm$. These are approximately the symmetric $\psi_{+}\approx(\psi_1(x+s/2,y)+\psi_1(x-s/2,y))/\sqrt{2}$ and the antisymmetric $\psi_{-}\approx(\psi_1(x+s/2,y)-\psi_1(x-s/2,y))/\sqrt{2}$ combinations of the single nanowire modes. Comparing this spectrum with the spectrum of the two site discrete model, we identify $t=(\beta_{+}-\beta_{-})/2$. In Fig.~\ref{fig:sym_antisym}(a) we plot $\beta_z$ of the symmetric and antisymmetric modes obtained using the Helmholtz equations as a function of $s$. We observe that the splitting of the symmetric and antisymmetric modes with respect to the single-nanowire $\beta_1$ (black dash line) is even for $s > 0.5 \, \mu \text{m}$ and hence we can extract the tight binding parameter $t$. For $s < 0.5 \, \mu \text{m}$ the splitting becomes uneven signaling the breakdown of the tight binding model. We plot the extracted tight-binding parameter $t=(\beta_{+}-\beta_{-})/2$, which is applicable for $s > 0.5 \, \mu \text{m}$, in Fig.~\ref{fig:sym_antisym}(b).

\begin{figure}[ht]
\includegraphics[width=3.4in]{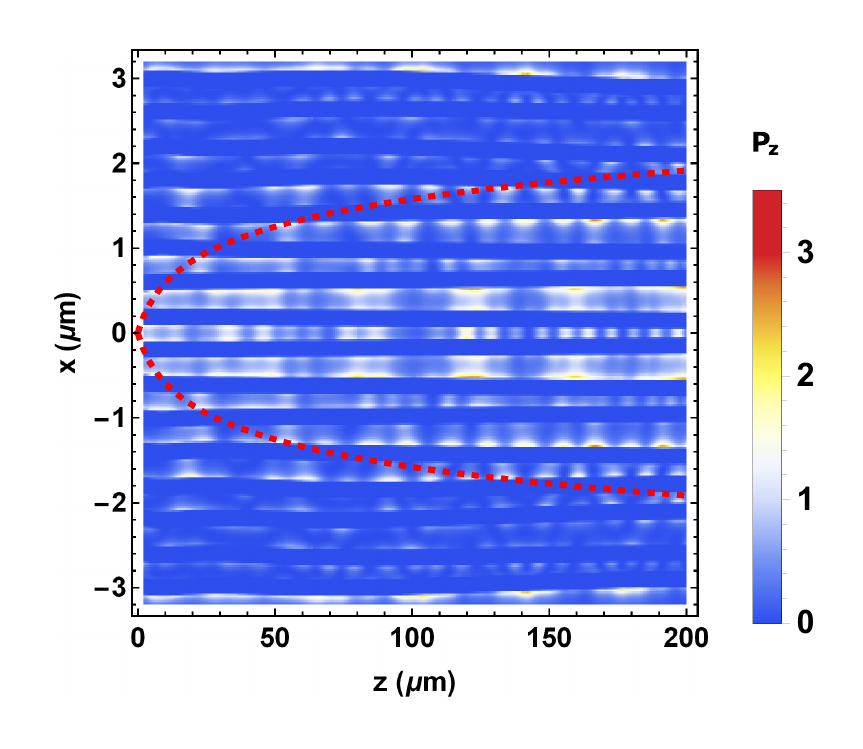}
\centering
\caption{\small{Mode rejection by the mode filtering nanowire array. On the input (left) side of the nanowire array, a bulk mode that is orthogonal to the two select modes, is injected into the nanowire array. On the output (right) side there is essentially no light in the vicinity of the two topological defect modes.}}
\label{fig:MFSup}
\end{figure}

\subsection{Mode filtering}
In Fig.~\ref{fig:MFSup} we demonstrate an example of mode rejection by the mode filter. We inject a mode that is not adiabatically connected to the two topological defect modes in the structure. The light flux spreads out over the whole structure except for the area in the vicinity of the two topological defects.

\subsection{Robustness}
\begin{figure*}[htbp]
\includegraphics[width=6.8in]{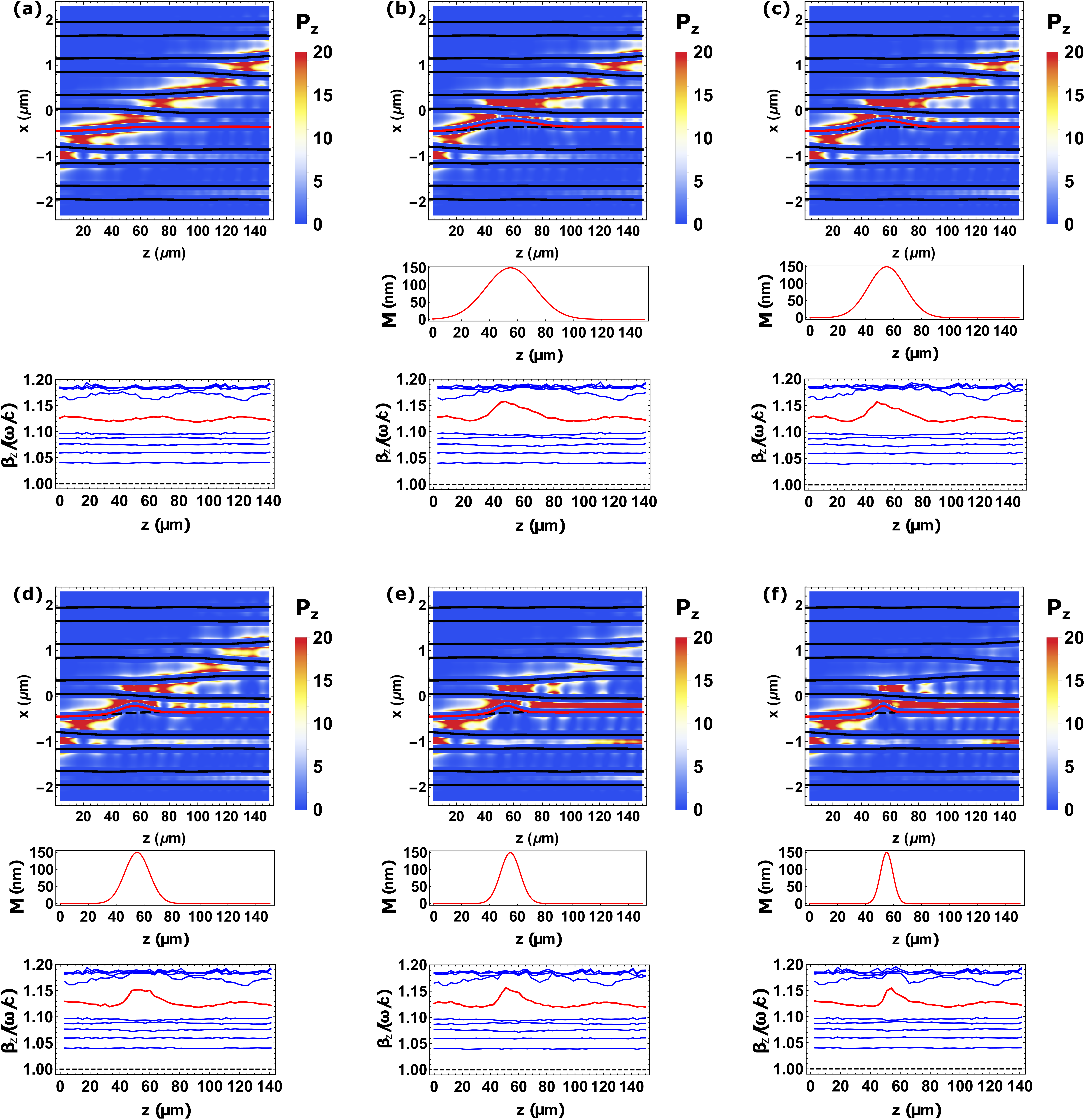}
\centering
\caption{\small{Light propagation along structures with a perturbation on the position of one of the nanowires. (a) Light flux (top panel) and the spectral flow (bottom panel) in a structure with no perturbation. (b)-(f) Light flux (top panel), meander $M(z)$ (middle panel), and spectral flow (bottom panel) in structures with one of the nanowires displaced from its original position (dashed line in top panel) to a new position (red line in top panel).  The nanowire is displaced by $\delta x=150$~nm over a width of $w=25.00 \, \mu$m (b), $18.75 \, \mu$m (c), $12.50 \, \mu$m (d), $10.00 \, \mu$m (e), $6.25 \, \mu$m (f).}}
\label{fig:sup1}
\end{figure*}

In this set of figures we provide additional data for the same set of structures that were used to construct Fig.~3 of the main text. The data demonstrate the robustness of the topological defect mode against  perturbation in nanowire position and diameter. The top panel of Fig~\ref{fig:sup1}(a) shows the light flux computed using 3D FDTD in a structure with a single topological defect shifting across the nanowire array. Light is injected into the topological defect mode on the left side of the array. The light is guided in the middle of the nanowire array and remains confined to the topological defect until it exits the array on the right side. The bottom panel of Fig~\ref{fig:sup1}(a) shows the spectral flow of the $\beta_z$ spectrum as a function of position along the wire (similar to Fig.~2(b) and (e) of the main text). The spectrum plot shows that the defect mode is well separated from the bulk modes throughout the structure. 

Next, we test robustness to perturbations by displacing the red nanowire by a Gaussian with maximum displacement of $\delta x$  and a width $w$. In Fig.~\ref{fig:sup1} (b) to (f) we plot the light flux, wire displacement, and the effective $\beta_z$ spectrum as a function of $z$ for $\delta x=150~\text{nm}$ and $w$ ranging from $25 \, \mu$m to $6.25 \, \mu$m. From the light flux plots (top panels), we observe that the light intensity is well guided when $w\gtrsim 12.5 \, \mu$m. For $w \lesssim 10.00 \, \mu$m, light flux heavily leaks into the bulk modes. The reason for this can be seen in the spectral flow plots. As the wire displacement becomes less smooth, the shift in the topological mode $\beta_z$ becomes more abrupt, and the adiabatic approximation fails. 

\begin{figure*}[htbp]
\includegraphics[width=6.8in]{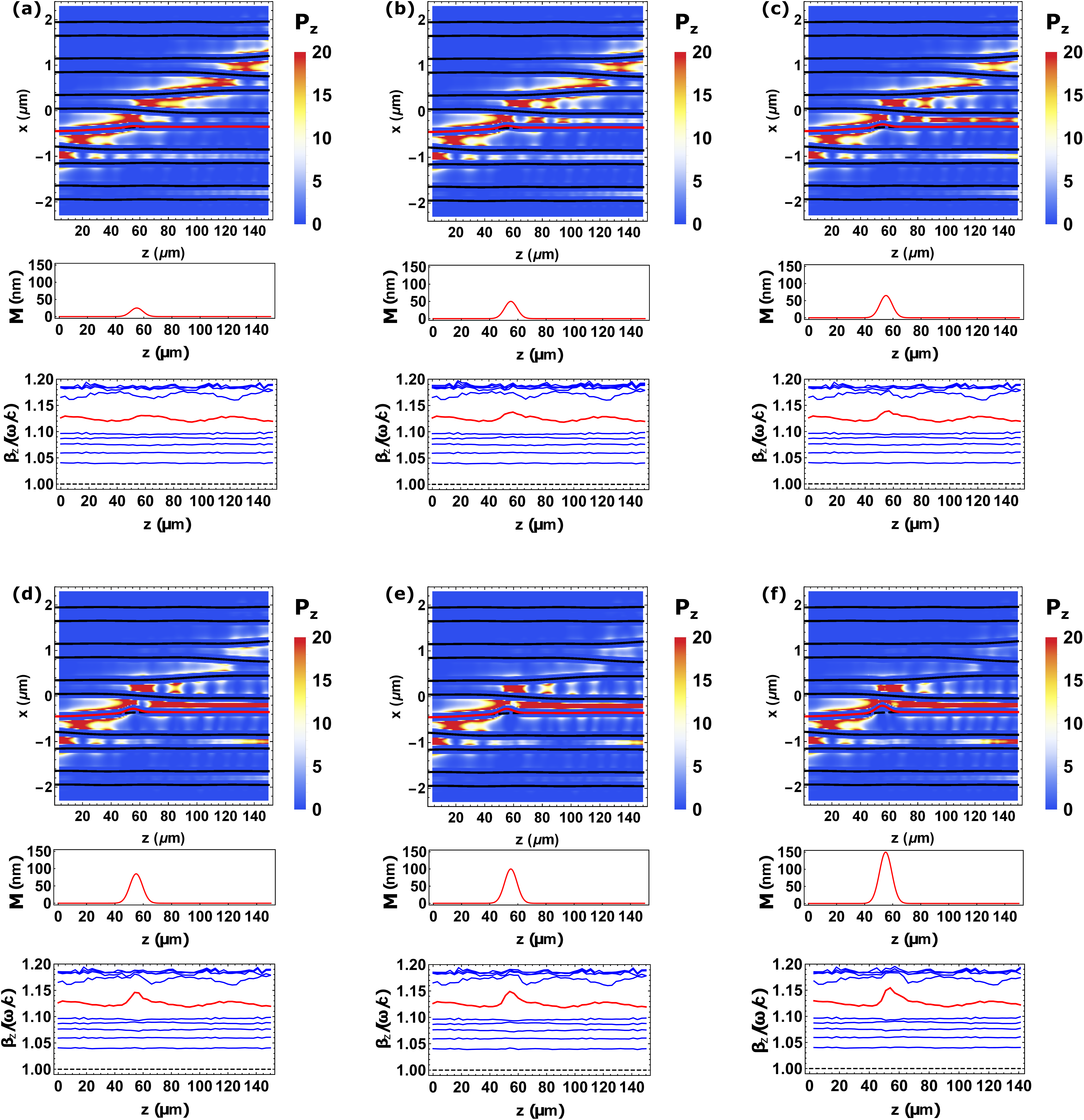}
\centering
\caption{\small{Same as Fig.~\ref{fig:sup1}, except we fix the displacement width at $w = 6.25 \mu$m, and vary the displacement amount $\delta x=25 \, \mu$m (a), $\delta x_\text{max}=50 \, \mu$m (b), $\delta x_\text{max}=65 \, \mu$m (c), $\delta x_\text{max}=85 \, \mu$m (d), $\delta x_\text{max}=100 \, \mu$m (e), $\delta x_\text{max}=150 \, \mu$m (f). }}
\label{fig:sup2}
\end{figure*}

Fig.~\ref{fig:sup2} is similar to Fig.~\ref{fig:sup1}, except we vary $\delta x$ from $25\,\text{nm}$ to $150\,\text{nm}$ while keeping $w$ fixed. The spectral flow calculation show that as $\delta x$ is increased, the topological mode $\beta_z$ approaches closer and closer to the band gap. Adiabaticity, and hence mode guidance, breaks down at $\delta x \sim 65$ nm - $85$ nm.

\begin{figure*}[htbp]
\includegraphics[width=6.8in]{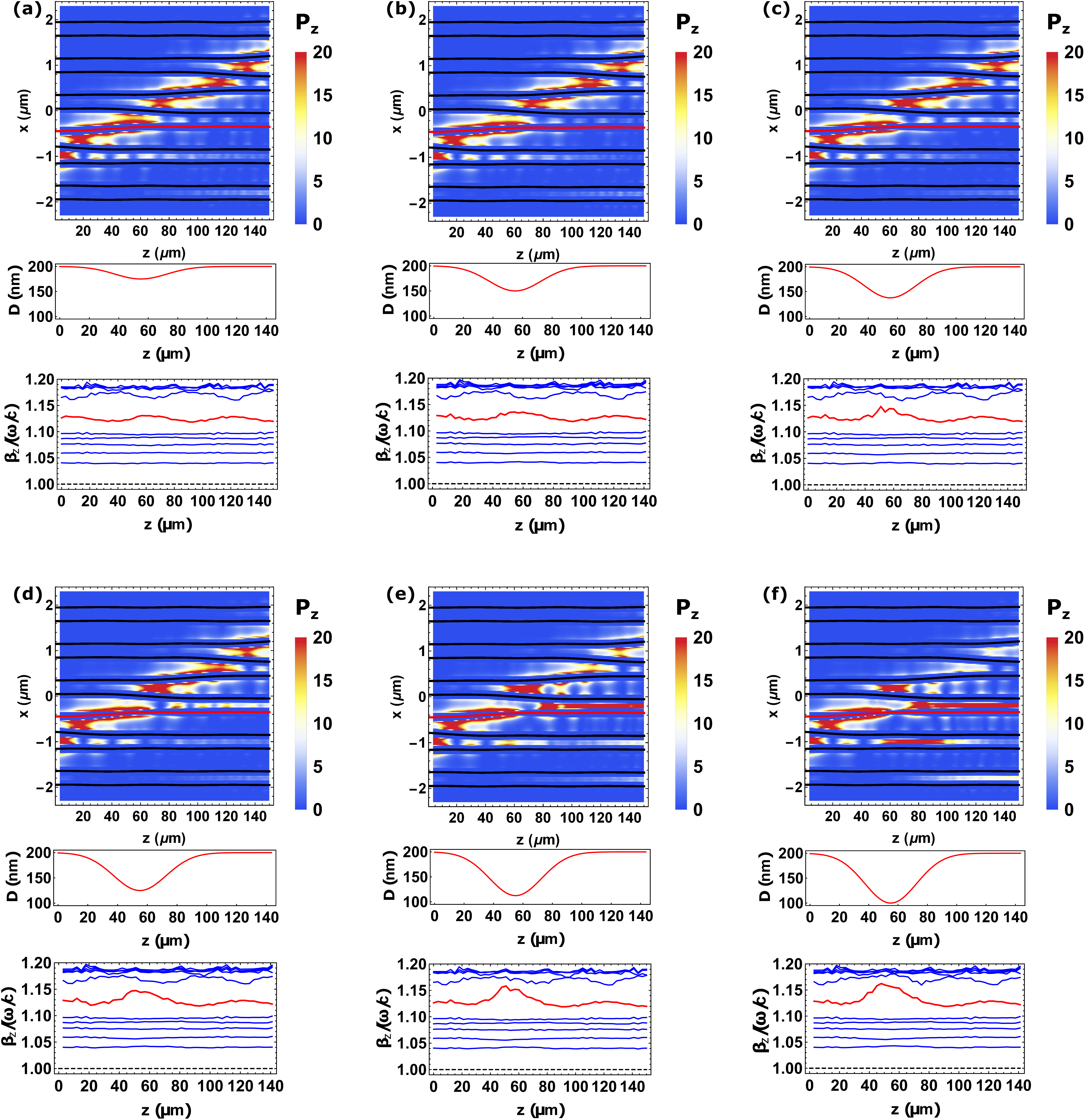}
\centering
\caption{\small{Same as Fig.~\ref{fig:sup1}, except we perturb the diameter $D (z)$ of the red nanowire. We fix the perturbation width at $w = 25 \mu$m, and shrink the nanowire diameter by $\delta D=25.0 \, \text{nm}$ (a), $\delta D=50.0 \, \text{nm}$ (b), $\delta D=62.5 \, \text{nm}$ (c), $\delta D=75.0 \, \text{nm}$ (d), $\delta D=87.5 \, \text{nm}$ (e), $\delta D=100 \, \text{nm}$ (f). }}
\label{fig:sup3}
\end{figure*}

Fig.~\ref{fig:sup3} is similar to Fig.~\ref{fig:sup1}, except we perturb the diameter $D$ of the red nanowire. The unperturbed nanowire has a diameter of $200$~nm, and we shrink it by $25.0\text{nm}$ to $100.0\,\text{nm}$ in panels (a)-(f). The perturbation profiles have a fixed width of $25 \,  \mu$m. Topological mode guidance starts to break down when the diameter shrinks by $87.5\,\text{nm}$ as the $\beta_z$ of the guided mode is approaching the bulk spectrum. 

\subsection{Decay length of defect mode in silver nanowire array}
\begin{figure}[ht]
\includegraphics[width=3.4in]{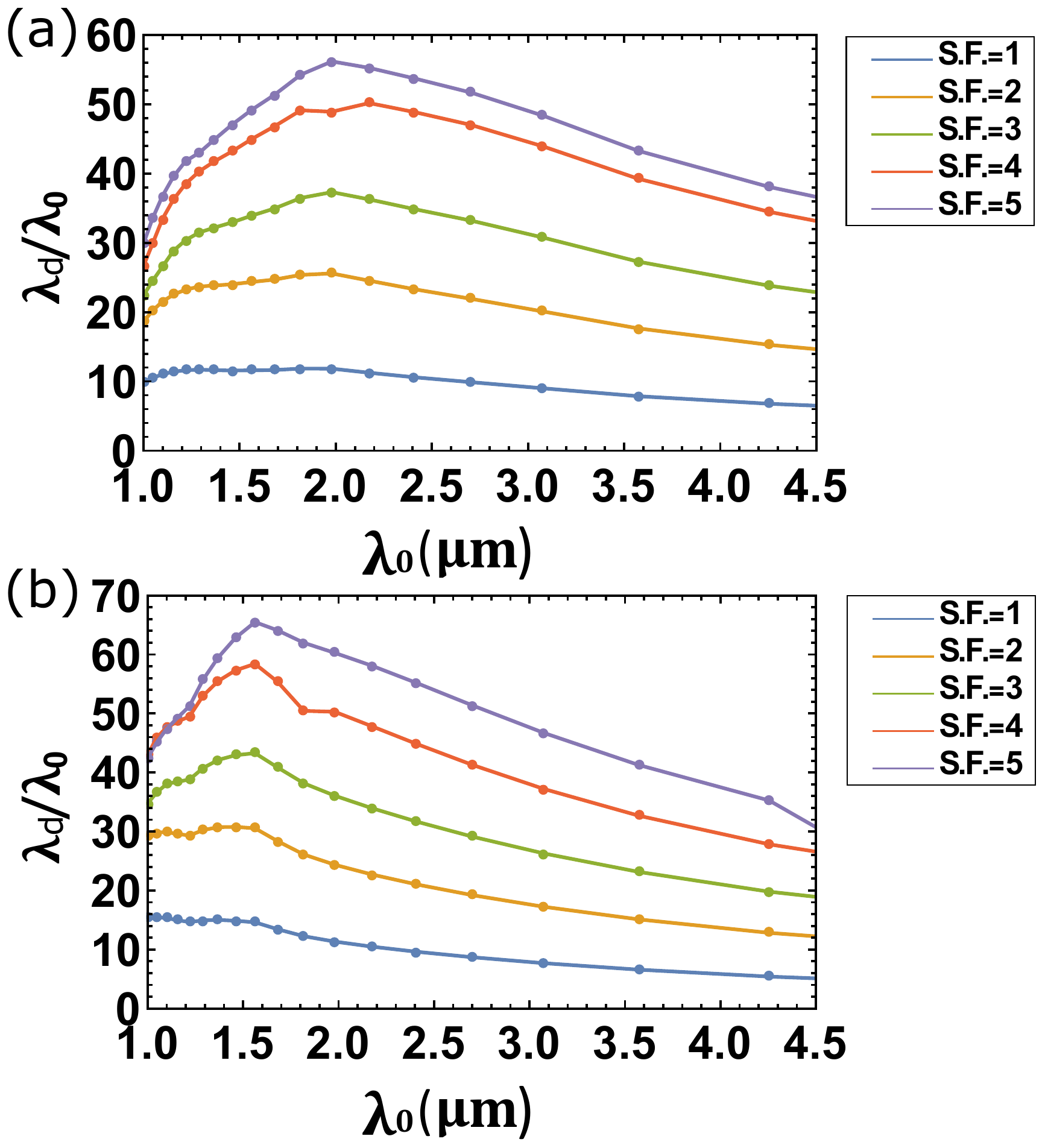}
\centering
\caption{\small{The ratio of the decay length of the topological defect mode to free space wavelength ($\lambda_\text{d} / \lambda_0 $) as a function of the free space wavelength ($\lambda_0$) for a silver nanowire array (a) and a gold nanowire array (b) with a single defect. The structure is enlarged by various scaling factors (S.F.) to enlarge the decay length of the topological defect mode.}}
\label{fig:decay}
\end{figure}

In this section, we compute the decay length of the topological defect modes in several silver and gold nanowire structures with a single topological defect. We use Lumerical Mode Solution numerical eigenmode solver to obtain complex $\beta_z$ and extract the decay length. The original nanowire array contains $15$ nanowires with $100$~nm radius which have staggered center-to-center spacings of $300$~nm and $500$~nm. In Fig.~\ref{fig:decay} we plot the decay length (normalized by $\lambda_0$) as a function of the wavelength in vacuum $\lambda_0$ for both silver and gold nanowires. We expect that for small $\lambda_0$ the decay length will be limited by the decrease in the optical conductivity of the metal while for large $\lambda_0$ the decay length will be limited by the penetration of light into the nanowires. Balancing these two effects we find that the largest ratio of decay length to $\lambda_0$ ($\lambda_\text{d} / \lambda_0 = 11.7$) in silver nanowires occurs for $\lambda_0=1.3 \, \mu$m. 

We can improve $\lambda_\text{d} / \lambda_0$ by going to longer $\lambda_0$ (to increase the optical conductivity of silver and gold) and enlarging the structure (to decrease the penetration of light into the nanowires). We therefore, scale the whole nanowire array by a scale factor (S.F.) $2$, $3$, $4$ and $5$, and plot the resulting $\lambda_\text{d} / \lambda_0$ as a function of $\lambda_0$ in Fig.~\ref{fig:decay}. We observe that as the nanowire array is enlarged, the $\lambda_0$ that corresponds to the peak of $\lambda_\text{d} / \lambda_0 $ ratio is red shifted, and the height of the peak of $\lambda_\text{d} / \lambda_0 $ increases. For scale factor (S.F.)  $5$ the peak $\lambda_\text{d} / \lambda_0$ ratio is $56.2$ at $\lambda_0=2.0 \, \mu$m for silver nanowires while $65.5$ at $\lambda_0=1.56 \, \mu$m for gold nanowires. 

\end{document}